\documentclass[english,showpacs,preprint,superscriptaddress]{revtex4-1}
\usepackage[T1]{fontenc}
\usepackage{babel}
\usepackage{textcomp}
\usepackage{amsmath}
\usepackage{graphicx}
\usepackage[unicode=true,
 bookmarks=true,bookmarksnumbered=true,bookmarksopen=true,bookmarksopenlevel=3,
 breaklinks=false,pdfborder={0 0 1},backref=false,colorlinks=false]
 {hyperref}
\hypersetup{
 linkcolor=black, citecolor=black, urlcolor=blue, filecolor=blue, pdfpagelayout=OneColumn, pdfnewwindow=true, pdfstartview=XYZ, plainpages=false}

\makeatletter

\providecommand{\tabularnewline}{\\}

\makeatother

\begin{document}

\title{Time-domain global similarity method for automatic data cleaning
for multi-channel measurement systems in magnetic confinement fusion
devices}

\author{Ting Lan}

\affiliation{School of Nuclear Science and Technology and Department of Modern
Physics, University of Science and Technology of China, Hefei, Anhui
230026, China}

\author{Jian Liu}
\email{Corresponding author: jliuphy@ustc.edu.cn}

\affiliation{School of Nuclear Science and Technology and Department of Modern
Physics, University of Science and Technology of China, Hefei, Anhui
230026, China}

\author{Hong Qin}

\address{School of Nuclear Science and Technology and Department of Modern
Physics, University of Science and Technology of China, Hefei, Anhui
230026, China}

\affiliation{Plasma Physics Laboratory, Princeton University, Princeton, NJ 08543,
USA}

\author{Lin Li Xu}

\affiliation{School of Computer Science and Technology, University of Science
and Technology of China, Hefei, Anhui 230026, China}
\begin{abstract}
{\large{}To guarantee the availability and reliability of data source
in Magnetic Confinement Fusion (MCF) devices, incorrect diagnostic
data, which cannot reflect real physical properties of measured objects,
should be sorted out before further analysis and study. Traditional
data sorting cannot meet the growing demand of MCF research because
of the low-efficiency, time-delay, and lack of objective criteria.
In this paper, a Time-Domain Global Similarity (TDGS) method based
on machine learning technologies is proposed for the automatic data
cleaning of MCF devices. Traditional data sorting aims to the classification
of original diagnostic data sequences, which are different in both
length and evolution properties under various discharge parameters.
Hence the classification criteria are affected by many discharge parameters
and vary shot by shot. The focus of TDGS method is turned to the physical
similarity between data sequences from different channels, which are
more essential and independent of discharge parameters. The complexity
arisen from real discharge parameters during data cleaning is avoided
in the TDGS method by transforming the general data sorting problem
into a binary classification problem about the physical similarity
between data sequences. As a demonstration of its application to multi-channel
measurement systems, the TDGS method is applied to the EAST POlarimeter-INterferomeTer
(POINT) system. The optimized performance of the method evaluated by 24-fold cross-validation has reached
0.9871\textpm 0.0385.}{\large \par}
\end{abstract}
\maketitle

\section{{\large{}Introduction}}

{\large{}With the development of Magnetic Confinement Fusion (MCF)
science and diagnostic techniques, massive diagnostic data are increasingly
generated. Original diagnostic data could be unreliable due to various
interference sources and complex measuring conditions in MCF devices,
such as mechanical vibration, electromagnetic interference, signal
saturation, and hardware failures. To guarantee the availability and
reliability of data source, incorrect diagnostic data, dubbed dirty
data, which cannot reflect real physical properties of measured objects,
should be sorted out before further analysis and study. The identification
of incorrect data can be regarded as a typical classification problem,
i.e., how to properly divide the original data set into two groups,
correct and incorrect one. Since experimental setups and discharge
processes are diverse, measured quantities from different shots, diagnostic
systems, and devices evolve in totally different ways. Incorrect diagnostic
results also vary due to their uncertain causes. Therefore, it is
difficult to define general and clear criteria for data cleaning.
Traditionally, dirty data are searched and removed manually with the
assistant of computer programs, mainly according to some simple rules,
common experiences, and sometimes personal intuitions. These data
cleaning programs and rules only apply to certain specific data and
usually performs poorly. Explosively increasing fusion data cannot
be satisfactorily cleaned in time. Real-time processing and feed-back
control require much faster data cleaning methods, which can remove
dirty data in a short time, say millisecond.
On the other hand, subjective factors in manual data cleaning processes
lead to inconsistent results. To meet the demand of fusion energy
research, the speed, efficiency, and accuracy of fusion data cleaning
should be improved imperatively. Automatic data cleaning methods based
on machine learning is a strong candidate for breaking through the
bottleneck of massive data application in fusion research.}{\large \par}

{\large{}In recent years, as computing ability and storage capacity
grow rapidly, Artificial Intelligence (AI) and machine learning have
been widely applied to a variety of scientific research fields, such
as image processing, biology, and astronomy\cite{dobkin1996computing,tarca2007machine,Way:2012:AML:2341297},
showing great advantages of extracting new patterns and principles
from complicated big data set. In MCF research, machine learning has
been applied in disruption prediction \cite{cannas2003disruption,ratta2010advanced,dormido2013development,cannas2010adaptive,yoshino2005neural,gerhardt2013detection},
plasma equilibrium parameters extraction \cite{lister1991fast}, data
retrieval \cite{vega2009overview}, L-H transition time estimation
\cite{vega2009automated}, charge exchange spectra analysis \cite{svensson1999analysis},
neoclassical transport database construction \cite{wakasa2007construction},
turbulent transport construction \cite{citrin2015real}, electron
temperature profile reconstruction \cite{clayton2013electron}, and
energy confinement scaling \cite{allen1992neural}. These pioneering
works have pushed the fusion energy research forward in many aspects
respectively. Wider, larger scale, and systematic application of machine
learning in fusion science will trigger radical changes. Effective
data cleaning also becomes the essential prerequisite for the application
of AI, data mining, and big data techniques in fusion research. Machine
learning in turn offers powerful tools for diagnostic data cleaning.
Precise data cleaning can be achieved by using objective classification
model trained by original data using supervised machine learning methods.
The application speed of well trained model will be easily optimized
to meet the requirement of real-time feedback control. With the support
of supercomputer, massive fusion data can be processed effectively
to relieve the data processing pressures of researchers. The robustness
and universality of classification models lays a foundation for the
large-scale applications of machine learning in fusion science.}{\large \par}

{\large{}In this paper, a new data cleaning method based on the Time-Domain
Global Similarity (TDGS) among data sequences defined by typical machine
learning technologies is proposed. The general-purposed TDGS method
can be used to automatically sort dirty diagnostic data from MUlti-channel
Measurement (MUM) systems in MCF experiments. Most diagnostic systems
of MCF devices are MUM systems, which measure the time evolution of
plasma parameters from different locations or directions with multiple
independent measuring channels, such as common interferometer systems
\cite{kawahata1999far}, polarimeter systems \cite{donne2004poloidal,brower2001multichannel,liu2014faraday,liu2016internal,zou2016optical},
electron cyclotron emission imaging systems \cite{luo2014quasi},
etc. Time sequences of diagnostic data from different channels of
the MUM system reflect related yet distinct aspects of the same observed
object. Therefore these diagnostic data are physically associated.
We define this relation as physical similarity. The physical similarity
just exists between correct data sequences from different channels
of the MUM system. The dirty data, which are caused by a variety of interference
sources, are physically dissimilar from correct data sequences or
each other. To overcome the difficulty of direct classification, the
TDGS method sorts the dirty data by classifying the physical similarity
between diagnostic data sequences from different channels under the
same discharge. Traditional data sorting aims to the classification
of original diagnostic data sequences, which are different in both
length and evolution properties under various discharge parameters.
Hence the classification criteria are affected by many discharge parameters
and vary shot by shot. The focus of TDGS method turns to the physical
similarity between data sequences from different channels, which are
more essential and independent of discharge parameters. Then the complexity
arisen from real discharge parameters during data cleaning is avoided
in TDGS method by transforming the general data sorting problem into
a binary classification problem about the physical similarity between
data sequences.}{\large \par}

{\large{}In TDGS method, the sample set is generated by the direct
sum of two original diagnostic data sequences from two different
channels of a MUM system in the same discharge. By combining two data sequences from different channels of an N-channel MUM system as one sample, $C_{N}^{2}$ samples can be generated for one discharge, and $P*C_{N}^{2}$ samples can be generated for $P$ discharges. Each sample is tagged by several indices
which indicates the corresponding physical similarity between two
sequences. These indices span a high dimensional index-space, in which
these samples can be classified into two groups, physical similar
samples and physical dissimilar ones. A physical similar sample is
constituted by two correct data sequences. If a sample is classified
to be physical dissimilar, its constituents contain at least one dirty
data sequence. According to this rule, the dirty diagnostic data can
be properly identified by physical similarity. In many MUM systems
of MCF devices, the physical similarity between diagnostic data exists
in time domain rather than frequency domain. And the dissimilarity
between dirty data, or dirty and clean data, is in global scale for
most of the time, instead of local and small scale, see Fig.\,\ref{fig:Line-averaged-density-measured}.
TDGS method employs different definitions of distance between two
time-series signals as tag indices of a sample, measuring this global
time-domain similarity. To guarantee precise classification, different
kinds of distance functions are adopted to map signals from a high-dimensional space of original data sequences into a lower-dimensional feature space. Because in different discharges the length of diagnostic
data sequences changes, original data sequences should be normalized
in distance functions to guarantee length independence. Then samples
generated from different discharges can be joined together as a unified
sample set, used as the training set or the test set. To demonstrate
the performance of TDGS method, it is tested by the data cleaning
of POlarimetry-INTerferometry (POINT) system on EAST Tokamak. The performance in this paper refers to the accuracy rate of classification results about physical similarity. In this
example, Support Vector Machine (SVM) is adopted as the classification
algorithm, which has advantage in solving non-linear, high-dimensional
problems \cite{cortes1995support,boser1992training,hsu2003practical,platt1998sequential}.
The k-fold cross-validation is used as model assessment method because
it can provide an effectively unbiased error estimate. In k-fold cross-validation
method, all samples contribute to both training and validation, and
each sample is used for validation only once \cite{kohavi1995study}.
For practical purpose of data cleaning for MCF devices, the samples
of a validation set are selected to be generated by the data from
one discharge, rather than randomly from the whole set. Evaluated
by 24-fold cross-validation, the accuracy rate of TDGS method in cleaning
the dirty density data of POINT system can reach 0.9871\textpm 0.0385,
which meets the application requirements of POINT system. By applying
TDGS method to the data cleaning of POINT system, the reliability
and availability of data source from POINT system are evidently enhanced
quickly and conveniently.}{\large \par}

{\large{}The rest parts of this paper are organized as follows. In
Sec. II, the theory and procedure of TDGS method for automatic diagnostic
data cleaning of MUM systems are introduced. In Sec. III, as an example,
TDGS method is applied to dirty data sorting of POINT system. In Sec.
IV, the assessment of TDGS method is studied in detail. In Sec. V,
the prospects of applying TDGS method to data cleaning in more MUM
systems of MCF devices are discussed. Moreover, the further optimization
of TDGS method is also proposed.}{\large \par}

\section{{\large{}Time-Domain Global Similarity method}}

{\large{}TDGS method aims to automatically sort out the incorrect
diagnostic data of the MUM system. By transforming the direct data
sequence classification problem into a binary classification problem
about the physical similarity between diagnostic data sequences, TDGS
method eliminates the complexity arisen from discharge parameters
during data cleaning. In this section, the theory and procedure of
training TDGS model for data cleaning of MUM systems are explained
in details, including data preprocessing, sample set generation, model
training, and application. }{\large \par}

{\large{}Preprocessing procedures of TDGS method contains digital
filtering and normalization. For many MUM systems in MCF devices,
the physical similarity between channels exists in global time scale.
Different filter methods can be used to remove the small time scale
fluctuations in original data. Here we choose the median filter method,
which is performed by letting a window move over the points of the
sequence and replacing the value at the center of the window by the
median of the original values within the window \cite{justusson1981median},
as digital filtering technique to eliminate the small-scale information.
The original sequence $\boldsymbol{S_{mj}}$ is transformed to $\boldsymbol{S'_{mj}}$
by median filter of window size $n$, namely}{\large \par}

{\large{}
\begin{equation}
\boldsymbol{S'_{mj}}=\mathrm{Median}Filter(\boldsymbol{S_{mj}},n),\label{eq:1}
\end{equation}
where $\boldsymbol{S_{mj}}$ denotes the original time series signal
measured by the $jth$ channel of the MUM system under the $mth$
discharge. By adaptively choosing the window size $n$ of median filter,
we efficiently remove high-frequency details, which matters little
to the global profile of data sequences.}{\large \par}

{\large{}The ranges of diagnostic data sequences from various channels,
discharge parameters, and diagnostic systems are different. To remove
the dependence of physical similarity on absolute values, TDGS method
normalizes all original data sequences to the same scale with Z-score
transformation. The sequence $\boldsymbol{S'_{mj}}$ is transformed
to $\boldsymbol{S''_{mj}}$ under Z-score transformation as}{\large \par}

{\large{}
\begin{equation}
\boldsymbol{S''_{mj}}=\frac{\boldsymbol{\boldsymbol{S'_{mj}}-E(\boldsymbol{S'_{mj}})}}{D(\boldsymbol{S'_{mj}})},\label{eq:2}
\end{equation}
where $E(\boldsymbol{S'_{mj}})$ and $D(\boldsymbol{S'_{mj}})$ denote
the average value and standard deviation of the sequence $\boldsymbol{S'_{mj}}$
respectively. After data preprocessing, the distinction of length
and magnitude of data sequences can be eliminated, which favors to
the generation of a unified sample set. }{\large \par}

{\large{}The second step of TDGS method is to define the sample 
$\boldsymbol{Sp_{ij}^{m}}$ , sample label $Label_{ij}^{m}$, and sample
feature $l_{ijk}^{m}$ using the preprocessed data. By combining any
pair of data sequences from different channels of the MUM system under
the same discharge, the sample set in TDGS method is defined as}{\large \par}

{\large{}
\begin{equation}
\boldsymbol{Sp_{ij}^{m}}=\boldsymbol{S''_{mi}}\oplus\boldsymbol{S''_{mj}},\label{eq:3}
\end{equation}
where the sample $\boldsymbol{Sp_{ij}^{m}}$ is the combination of
preprocessed data sequence $\boldsymbol{S''_{mi}}$ from the $i$th
channel and $\boldsymbol{S''_{mj}}$ from the $j$th channel of the
MUM system under the $mth$ discharge. The sample label $Label_{ij}^{m}$ assigned to the sample $\boldsymbol{Sp_{ij}^{m}}$ is used to judge the corresponding physical similarity between the preprocessed data
$\boldsymbol{S''_{mi}}$ and $\boldsymbol{S''_{mj}}$. The physical similarity is labeled according to the correctness of data sequences. If the data sequences from channel $i$ and channel $j$ under shot $m$ are correct, the label $Label_{ij}^{m}$ is tagged as $+1$, which means these two sequences are physical similar with each other. If the data sequences from channel $i$ and channel $j$ under shot $m$ contain at least one incorrect data sequence, the label $Label_{ij}^{m}$  is tagged as $-1$, which means these two sequences are physical dissimilar with each other. From the preprocessed
data of an N-channel MUM system under $P$ discharges, $P*C_{N}^{2}=\frac{P*N(N-1)}{2}$
samples can be generated. }{\large \par}

{\large{}In many MUM systems of MCF devices, there exists physical
similarity between diagnostic data sequences in time domain. Considering
that the dimension of the sample is too large for input, say a million,
reducing the dimension by selecting a set of principal features is
necessary to avoid the curse of dimension and relieve the heavy calculation
burden. TDGS method employs multiple distance functions between data
sequences as sample features, i.e.,}{\large \par}

{\large{}
\begin{equation}
l_{ijk}^{m}=D_{k}(\boldsymbol{S''_{mi}},\boldsymbol{S''_{mj}}),\label{eq:4}
\end{equation}
where $D_{k}$ is the $kth$ distance function employed in TDGS method,
and sample feature $l_{ijk}^{m}$ denotes the $kth$ sample feature
extracted from the sample $\boldsymbol{Sp_{ij}^{m}}$. The definitions
of all the distance functions $D_{k}$ are listed in Table.\,\ref{table_feature}.
A variety of distance functions, such as Euclidean Distance, Chebyshev
Distance, and Correlation distance, are selected for measuring the
physical similarity in various aspects. The contribution of Chebyshev
Distance is remarkable in the classification of the sudden change
at some point in the data sequence. The first-order and second-order
differential of Euclidean Distance are more important in the case
when the variation tendency of incorrect data sequences is obviously
different from the correct ones. Some distance functions are affected by lengths of data sequences. For example, the Euclidean Distance would be larger if lengths of input signals are longer. The lengths of data sequences would be longer if the time durations of corresponding discharges are longer or the sampling rates are higher. But the sample features should not be affected by the time duration of discharges or the sampling rates of signals. So those distance functions which are affected by lengths of corresponding data sequences are normalized by the length $\boldsymbol{T}$ of data sequences. Then samples from different discharges with various time
durations and sampling rates can be normalized to the same standard for similarity.}{\large \par}

{\large{}}
\begin{table}[h]
{\large{}\caption{Mathematical definitions of distance functions as 11 features of a
sample adopted in TDGS method\label{table_feature}. $\boldsymbol{T}$ is the length of the input data sequence. }
}{\large \par}
\centering{}{\large{}}%
\begin{tabular}{|c|c|}
\hline 
{\large{}Feature id} & {\large{}$D_{k}(\boldsymbol{S_{1},S_{2}})$)}\tabularnewline
\hline 
\hline 
{\large{}1} & {\large{}$\frac{\Vert\boldsymbol{S_{1}-S_{2}}\Vert_{2}}{\boldsymbol{\boldsymbol{T}}}$}\tabularnewline
\hline 
{\large{}2} & {\large{}$\frac{\Vert\boldsymbol{S_{1}-S_{2}}\Vert_{1}}{\boldsymbol{\boldsymbol{T}}}$}\tabularnewline
\hline 
{\large{}3} & {\large{}$\Vert\boldsymbol{S_{1}-S_{2}}\Vert_{\infty}$}\tabularnewline
\hline 
{\large{}4} & {\large{}$\ensuremath{\frac{{{{\bf {S}}_{1}}^{T}{{\bf {S}}_{2}}}}{{\parallel{{\bf {S}}_{1}}{\parallel_{2}}\parallel{{\bf {S}}_{2}}{\parallel_{2}}}}}$}\tabularnewline
\hline 
{\large{}5} & {\large{}$\ensuremath{\frac{{E(({{\bf {S}}_{1}}-E({{\bf {S}}_{1}}))({{\bf {S}}_{2}}-E({{\bf {S}}_{2}})))}}{{\sqrt{D({{\bf {S}}_{1}})}\sqrt{D({{\bf {S}}_{2}})}}}}$}\tabularnewline
\hline 
{\large{}6} & {\large{}$\ensuremath{1-{D_{5}}({{\bf {S}}_{{\bf {1}}}},{{\bf {S}}_{2}})}$}\tabularnewline
\hline 
{\large{}7} & {\large{}$\ensuremath{{{\parallel{{d\left({{{\bf {S}}_{1}}-{{\bf {S}}_{2}}}\right)}\mathord{\left/{\vphantom{{d\left({{{\bf {S}}_{1}}-{{\bf {S}}_{2}}}\right)}{dt}}}\right.\kern -\nulldelimiterspace}{dt}}{\parallel_{2}}}\mathord{\left/{\vphantom{{\parallel{{d\left({{{\bf {S}}_{1}}-{{\bf {S}}_{2}}}\right)}\mathord{\left/{\vphantom{{d\left({{{\bf {S}}_{1}}-{{\bf {S}}_{2}}}\right)}{dt}}}\right.\kern -\nulldelimiterspace}{dt}}{\parallel_{2}}}{\bf {T}}}}\right.\kern -\nulldelimiterspace}{\bf {T}}}}$}\tabularnewline
\hline 
{\large{}8} & {\large{}$\ensuremath{{{\parallel{{d\left({{{\bf {S}}_{1}}-{{\bf {S}}_{2}}}\right)}\mathord{\left/{\vphantom{{d\left({{{\bf {S}}_{1}}-{{\bf {S}}_{2}}}\right)}{dt}}}\right.\kern -\nulldelimiterspace}{dt}}{\parallel_{1}}}\mathord{\left/{\vphantom{{\parallel{{d\left({{{\bf {S}}_{1}}-{{\bf {S}}_{2}}}\right)}\mathord{\left/{\vphantom{{d\left({{{\bf {S}}_{1}}-{{\bf {S}}_{2}}}\right)}{dt}}}\right.\kern -\nulldelimiterspace}{dt}}{\parallel_{1}}}{\bf {T}}}}\right.\kern -\nulldelimiterspace}{\bf {T}}}}$}\tabularnewline
\hline 
{\large{}9} & {\large{}$\ensuremath{{{\parallel{{{d^{2}}\left({{{\bf {S}}_{1}}-{{\bf {S}}_{2}}}\right)}\mathord{\left/{\vphantom{{{d^{2}}\left({{{\bf {S}}_{1}}-{{\bf {S}}_{2}}}\right)}{d{t^{2}}}}}\right.\kern -\nulldelimiterspace}{d{t^{2}}}}{\parallel_{2}}}\mathord{\left/{\vphantom{{\parallel{{{d^{2}}\left({{{\bf {S}}_{1}}-{{\bf {S}}_{2}}}\right)}\mathord{\left/{\vphantom{{{d^{2}}\left({{{\bf {S}}_{1}}-{{\bf {S}}_{2}}}\right)}{d{t^{2}}}}}\right.\kern -\nulldelimiterspace}{d{t^{2}}}}{\parallel_{2}}}{\bf {T}}}}\right.\kern -\nulldelimiterspace}{\bf {T}}}}$}\tabularnewline
\hline 
{\large{}10} & {\large{}$\ensuremath{{{\parallel{{{d^{2}}\left({{{\bf {S}}_{1}}-{{\bf {S}}_{2}}}\right)}\mathord{\left/{\vphantom{{{d^{2}}\left({{{\bf {S}}_{1}}-{{\bf {S}}_{2}}}\right)}{d{t^{2}}}}}\right.\kern -\nulldelimiterspace}{d{t^{2}}}}{\parallel_{1}}}\mathord{\left/{\vphantom{{\parallel{{{d^{2}}\left({{{\bf {S}}_{1}}-{{\bf {S}}_{2}}}\right)}\mathord{\left/{\vphantom{{{d^{2}}\left({{{\bf {S}}_{1}}-{{\bf {S}}_{2}}}\right)}{d{t^{2}}}}}\right.\kern -\nulldelimiterspace}{d{t^{2}}}}{\parallel_{1}}}{\bf {T}}}}\right.\kern -\nulldelimiterspace}{\bf {T}}}}$}\tabularnewline
\hline 
{\large{}11} & {\large{}$\ensuremath{\mathop{(\mathop\sum\limits ^{10}}\limits _{k=1}{D_{k}}{({{\bf {S}}_{{\bf {1}}}},{{\bf {S}}_{2}})^{2}}{)^{0.5}}}$}\tabularnewline
\hline 
\end{tabular}{\large \par}
\end{table}
{\large \par}

{\large{}After magnitude normalization and time normalization, the
scale of data sequences is unified. Then samples from different discharges
can be fairly treated in the sample set $(\boldsymbol{X,Y})$, defined
by its components}{\large \par}

{\large{}
\begin{eqnarray}
\begin{cases}
\boldsymbol{X=}\left\{ \boldsymbol{\underset{k}{\oplus}}l_{ijk}^{m}\right\} _{i,j,m},\\
\boldsymbol{Y=\left\{ Label_{ij}^{m}\right\} }_{i,j,m}.
\end{cases}\label{eq:5}
\end{eqnarray}
}{\large \par}

{\large{}In the training procedures of TDGS method, many classification
algorithms can be used. Here we use Support Vector Machine (SVM) to
train classifiers for the physical similarity between data sequences.
In SVM, input samples are mapped to a high-dimensional feature space.
A good classification is achieved by constructing a linear separating
hyperplane in this feature space with the maximal margin to the nearest
samples of any class \cite{cortes1995support,boser1992training,hsu2003practical}.
Mathematically, the SVM seeks the solution of the following optimization
problem}{\large \par}

{\large{}
\begin{equation}
\ensuremath{\left\{ \begin{array}{l}
\mathop{\min}\limits _{{\bf {W}},{\kern1pt }{\kern1pt }b,{\kern1pt }{\kern1pt }{\bf {\xi}}}\frac{1}{2}{{\bf {W}}^{T}}{\bf {W}}+C\sum\limits _{i=1}^{l}{\xi_{i}},\\
subject{\kern1pt }{\kern1pt }{\kern1pt }to{\kern1pt }{\kern1pt }{\kern1pt }{\kern1pt }{{\bf {Y}}_{i}}({{\bf {W}}^{T}}\phi({{\bf {X}}_{i}})+b)\ge1-{\xi_{i}},\\
{\xi_{i}}\ge0,
\end{array}\right.}\label{eq:6}
\end{equation}
where $\boldsymbol{W}$ is the normal vector of the targeted separating
hyperplane, $C$ is the penalty parameter of the error term. And $\ensuremath{K({{\bf {X}}_{i}},{\kern1pt }{\kern1pt }{\kern1pt }{{\bf {X}}_{j}})=\phi{({{\bf {X}}_{i}})^{T}}\phi({{\bf {X}}_{j}})}$
is defined as the kernel function. Proper selection of kernel function
for different classification problems can optimize the performance
by mapping samples to appropriate high-dimensional feature space.
Sequential minimal optimization (SMO) is adopted as a common iterative
method for solving this quadratic programming (QP) problem \cite{platt1998sequential}. }{\large \par}

{\large{}After classification of the physical similarity, dirty diagnostic
data of MUM systems can be identified according to the tagged similarity
relations between data sequences. On the one hand, by scanning through
all samples tagged with similarity, the data which are similar to
each other under the same discharge can be marked as correct data.
On the other hand, by scanning through all samples tagged with dissimilarity,
the data which are dissimilar from all the other diagnostic data sequences
under the same discharge can be tagged as incorrect data. Calculation
burden and inconsistent error accumulation of data cleaning depends
on the judgment rules given the physical similarity classification
has been finished. There are still some optimization schemes for improving
the performance of data cleaning at this stage. The rule adopted in this paper is an absolute judging rule. Next step, we would adopt a non-absolute judging rule. For example, the sequence which is dissimilar from $90\%$ of the other sequences can be tagged as incorrect data. Then the degree parameter introduced by the judging rule can change the mapping relations between performance of TDGS method about physical similarity and correctness of data sequences. In some cases, proper setting of the degree parameter would improve the data cleaning performance of TDGS method.}{\large \par}

\section{{\large{}Applications of TDGS method on density data cleaning for
polarimetry-interferometry system}}

{\large{}In this section, TDGS method is used to clean the density
data of POINT system as an application example. By applying TDGS method
to sort out the incorrect density data automatically, the reliability
and availability of diagnostic data measured by POINT system can be
improved, which is beneficial for the study and application of POINT
data, for example, maintaining steady-state plasma. }{\large \par}

\begin{figure}[h]
{\large{}\caption{Original time-series density sequences from different channels of
EAST POlarimeter-INterferomeTer (POINT) system are plotted in three
typical discharges (shot 62287, 56180, 58888). The channel id of corresponding
sequences is labeled. The incorrect sequences are marked with red
boxes.\label{fig:Line-averaged-density-measured}}
}{\large \par}
\centering{}{\large{}\includegraphics[scale=0.3]{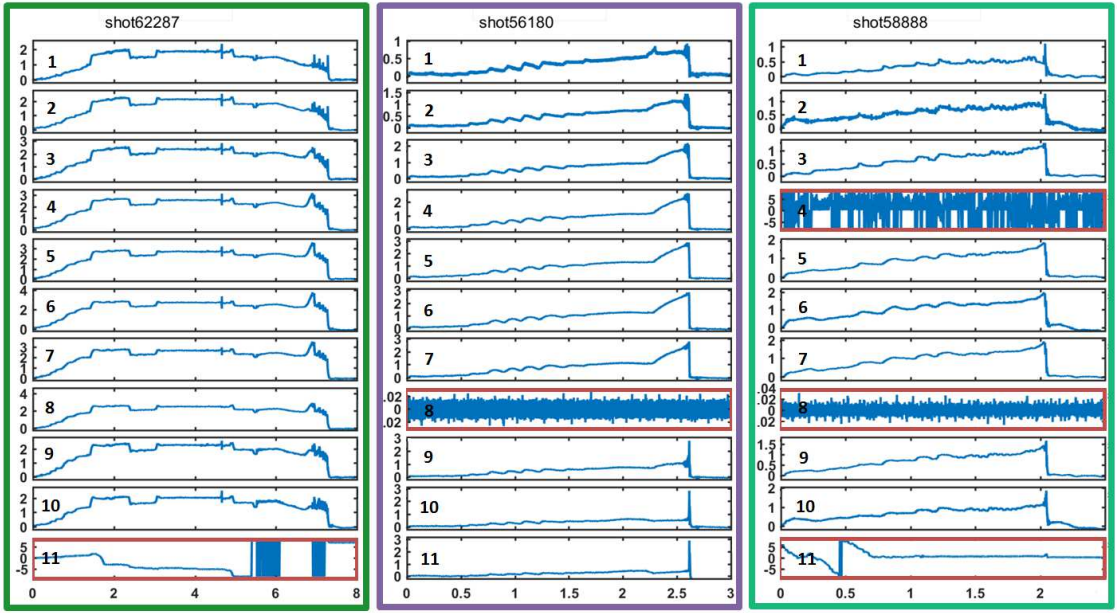}}{\large \par}
\end{figure}

{\large{}POINT is a typical 11-channel MUM system, which measures
line-average electron density of EAST tokamak at different vertical
locations with independent measuring channels. \cite{liu2014faraday,liu2016internal,zou2016optical}.
Original density data of POINT system contains dirty data sequences
caused by mechanical vibration, electromagnetic interference, collinear
error, or hardware failures \cite{zou2016optical}. Original density
sequences of POINT system under three discharges with different parameters,
i.e., shot 62287, 56180, and 58888, are plotted in Fig.\,\ref{fig:Line-averaged-density-measured}.
By comparing any two time evolutions of density from different shots,
it can be observed that the evolutions of density sequences under various
discharges are totally different from each other. Meanwhile, the incorrect
density sequences evolve differently even under the same discharge,
see the three incorrect data sequences marked with red boxes in shot
58888. It is difficult to find direct, general, and clear criteria
for sorting out the incorrect density sequences even only for the
three shots shown in Fig.\,\ref{fig:Line-averaged-density-measured}.
By further observation, the correct density data of different channels
under the same charge are globally similar to each other. For example,
all evolutions of density from channel 1 to channel 10 in shot 62287
can be recognized to have three stages, i.e., the climbing stage from
0 s to 1.5 s, the plateau stage from 1.5 s to 7 s, and the declining
stage from 7 s to 7.3 s. Correct data sequences from different channels of POINT system reflect plasma properties with independent measurement channels at different locations. There exits time-domain global similarity in correct density signals of POINT system. This similarity originates from the associated relation between plasma properties at different locations. Based on this physical similarity, TDGS method
can be efficient to sort out the incorrect density data of POINT system. }{\large \par}

{\large{}Data preprocessing of this application contains digital filtering
and normalization. There exits local differences between correct time-series
density sequences under the same charge. For example, the data sequences
from channel 1 to channel 10 in shot 62287 evolve differently at the
start moment of the declining stage, see Fig.\,\ref{fig:Line-averaged-density-measured}.
Removing these small-scale fluctuations by median filter can improve
the classification performance for physical similarity. Meanwhile,
the ranges and lengths of time-series density sequences under various
discharges are different. To guarantee the scale independence on absolute values, original
density sequences should be normalized to the same magnitude scale with Z-score
transformation. To guarantee the length independence, feature 1, 2,
7, 8, 9, 10, see Table.\,\ref{table_feature}, should be normalized
with the length of sequences. Considering that the sampling rate of
POINT system is unchanged during the selected 24 shots, we adopt the
acquisition time of corresponding discharges directly as the indices
for length normalization. After scale normalization and length normalization,
the effect of absolute value and length of sequences can be eliminated.
Then a unified sample set can be generated from the preprocessed data.}{\large \par}

{\large{}}
{\large \par}

\begin{figure}[h]
{\large{}\caption{In the sample set generated from density data of POINT system, the
ratios of bad channels and dissimilar samples are plotted for each
shot. \label{fig:In-the-labeled}}
}{\large \par}
\centering{}{\large{}\includegraphics[scale=0.4]{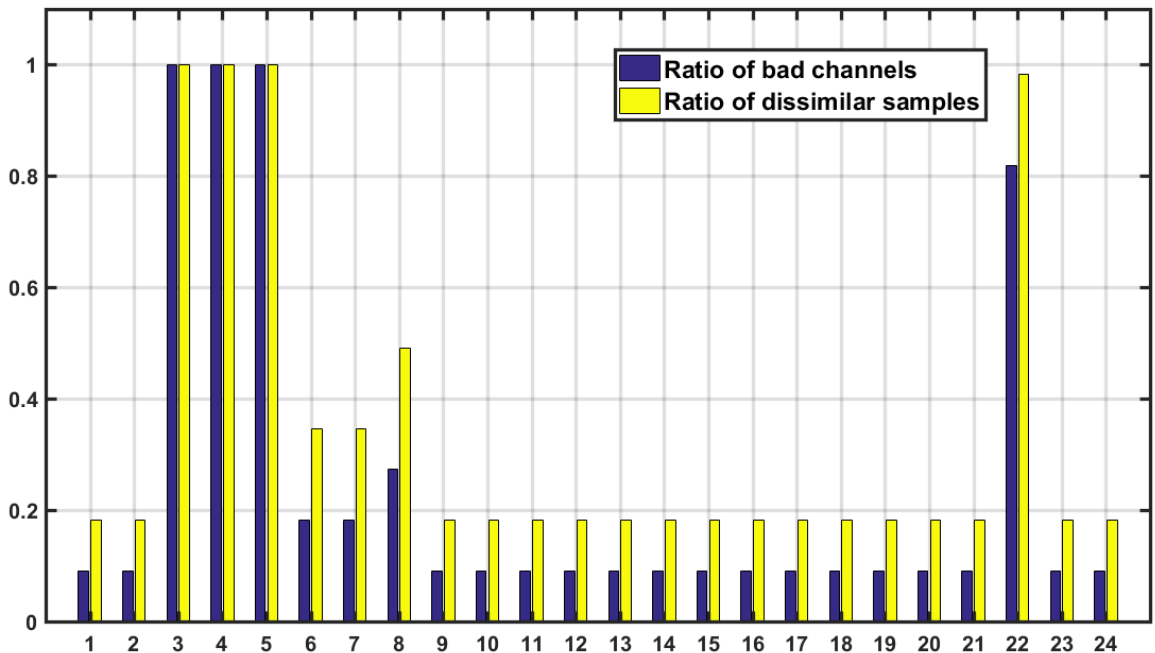}}{\large \par}
\end{figure}

{\large{}In conventional operations of POINT system, the ratio of
incorrect density data is much less than the ratio of correct ones.
When samples of one class are much more than the other class, most
classifiers are biased towards the major class and lead to very poor
classification rates on minor class. To improve the classification
accuracy on cleaning of density data for POINT system, we balance
the training database by undersampling the majority class, i.e., the
correct density data. To reveal the performance of the classification
model in general sense, different types of incorrect density data
of POINT system are collected in the training set as far as possible.
In this application, density data measured by 11 channels of POINT
system under 24 discharges are chosen to generate the sample set.
Considering that each sample of TDGS method is produced by combining
two diagnostic data sequences from different channels under the same
discharge, $24*C_{11}^{2}=1320$ samples can be generated for this
11-channel MUM system. To demonstrate the distribution structure of
data set, the ratios of bad channels and dissimilar samples for each
single discharge are separately shown in Fig.\,\ref{fig:In-the-labeled}.
Here the ratio of bad channels denotes the proportion of incorrect
diagnostic data sequences to total diagnostic data sequences. And
the ratio of dissimilar samples denotes the proportion of samples
tagged with dissimilarity to total samples. According to the definition
of physical similarity in TDGS method, two types of samples are tagged
with dissimilarity in this application. Type A is consist of a correct
density data sequence and an incorrect one under the same discharge.
Type B is consist of two incorrect density data sequences from different
channels under the same discharge. For example, the ratio of bad channels
of POINT system for shot 58888 is 3/11, see Fig.\,\ref{fig:Line-averaged-density-measured}.
Corresponding numbers of dissimilar samples of Type A and Type B are
$C_{8}^{1}*C_{3}^{1}=24$ and $C_{3}^{2}=3$ respectively. Hence the
ratio of dissimilar samples in shot 58888 is $(C_{8}^{1}*C_{3}^{1}+C_{3}^{2})/C_{11}^{2}=$27/55.
The ratio of total similar samples to total dissimilar samples in
the sample set is about 1.9. Different classes are almost balanced
in the training set, which is good for obtaining a better prediction
performance using TDGS method.}{\large \par}

{\large{}}
{\large \par}

{\large{}Different classification models with various parameter settings
can be adopted in TDGS method. By testing the performance of candidate
models, most suitable classifier with optimized parameters will be
selected for real applications. In this case, SVM is chosen as the
classification algorithm. SMO is adopted as the iterative
method. Key parameters, including the window size
of median filter, kernel functions of SVM, and penalty factor are
optimized by model selection. By comparing the performance of candidate models, most suitable classifier with optimized parameters will be selected. Proper window size of median filter can eliminate the  small-scale
information while preserving the feature of physical similarity for
plasma density. Appropriate penalty factor helps to avoid overfitting.
Linear, Polynomial, and Gaussian radial basis (RBF) kernel functions
are applied in this problem, see Table.\,\ref{tab:Basic-kernel-functions-1}.
Linear kernel is the simplest kernel function among them, with least
adjustable parameters. RBF kernel has an additional parameter $\gamma$,
while the Polynomial kernel has more adjustable parameters, including
the slope $\alpha$, the constant term $r$, and the polynomial order
$d$. To reduce the calculation amount during the parameter optimization
of Polynomial kernel, the value of polynomial order $d$ is restricted
to 3 and 4. By selecting a kernel function and corresponding kernel
parameters, the classification can be handled in a proper high-dimensional
feature space. Following these operation procedures, the optimized
classification model for density data cleaning of POINT system can
be achieved by TDGS method.}{\large \par}

{\large{}}
\begin{table}[h]
{\large{}\caption{Kernel functions chosen for optimization in the application of TDGS
method on density data cleaning for POINT system. {\small{}Here $\alpha$,
$r$, $d$, and $\gamma$ are corresponding kernel parameters. }\label{tab:Basic-kernel-functions-1}}
}{\large \par}
\centering{}{\large{}}%
\begin{tabular}{|c|c|}
\hline 
{\large{}Kernel function} & {\large{}Mathematical form}\tabularnewline
\hline 
\hline 
{\large{}Linear} & {\large{}$\ensuremath{{\bf {X}}_{i}^{T}{{\bf {X}}_{j}}}$}\tabularnewline
\hline 
{\large{}Polynomial} & {\large{}$\ensuremath{{(\alpha{\bf {X}}_{i}^{T}{{\bf {X}}_{j}}+r)^{d}},\alpha>0}$}\tabularnewline
\hline 
{\large{}Radial basis function (RBF)} & {\large{}$\ensuremath{\exp(-\gamma\parallel{{\bf {X}}_{i}}-{{\bf {X}}_{j}}{\parallel^{2}}),\gamma>0}$}\tabularnewline
\hline 
\end{tabular}{\large \par}
\end{table}
{\large \par}

\section{{\large{}Assessment of TDGS method}}

{\large{}In this section, model selection and assessment of TDGS method
are introduced. Corresponding assessment results of applying TDGS
method to density data cleaning for POINT system are exhibited, which
demonstrates the advantage of applying TDGS method to MUM systems
in MCF devices.}{\large \par}

\begin{figure}[h]
{\large{}\caption{In each test of 24-fold cross-validation, the classification accuracy
of applying TDGS method to the sorting of density data measured by
POINT system with various kernel functions is plotted. Here the legend
\textquoteleft linear\textquoteright{} denotes linear kernel; \textquoteleft polynomial
(3)\textquoteright{} denotes $3^{rd}$-order polynomial kernel; \textquoteleft polynomial
(4)\textquoteright{} denotes $4^{rd}$-order polynomial kernel; \textquoteleft rbf\textquoteright{}
denotes gaussian radial basis kernel function, and the scaling factor
(sigma) equals 1.\label{fig:In-every-test}}
}{\large \par}
\centering{}{\large{}\includegraphics[scale=0.4]{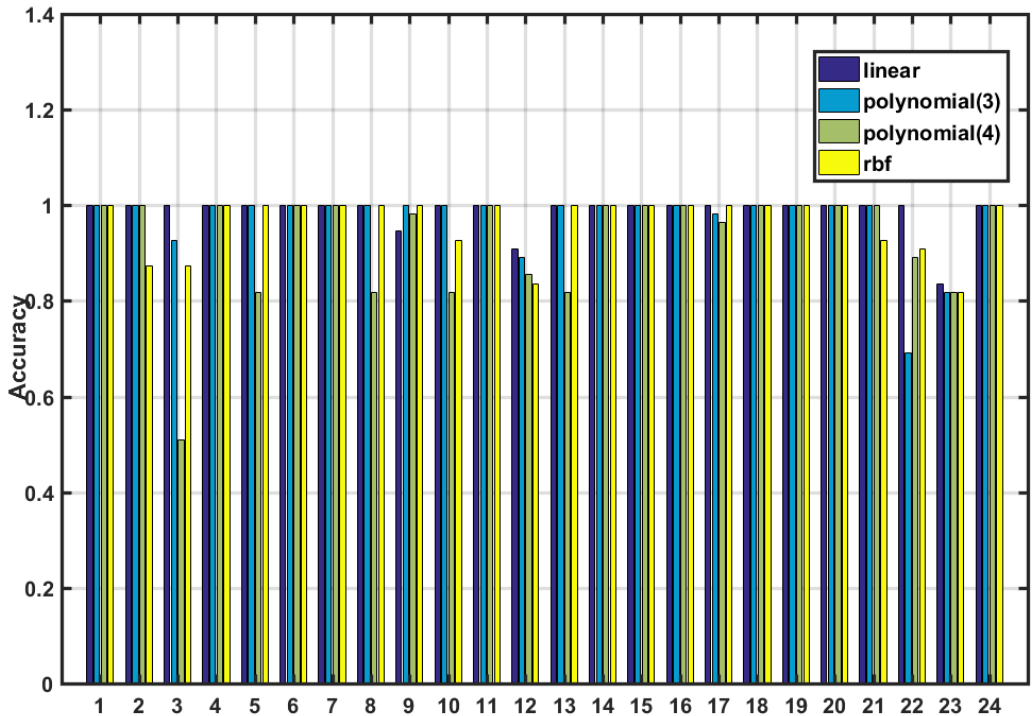}}{\large \par}
\end{figure}

{\large{}To provide an effectively unbiased error estimate, k-fold
cross-validation is adopted as the assessment method for TDGS method.
In k-fold cross-validation, the sample set is evenly divided into
k groups. One group is chosen as the validation set, and the rest
k-1 groups constitute the training set. Then, the assessment process
is repeated k times by assigning each group as the validation set
once \cite{browne2000cross}. In the data cleaning for MCF devices
during experimental period, the test sample set is generated from
the same discharge. To assessment the performance of TDGS method for
practical application, we govern that the samples in each group are
generated from the same discharge. For the sample set generated from
density data under 24 discharges, the model assessment method of POINT
system turns out to be 24-fold cross-validation. In each test of the
24-fold cross-validation, the classification accuracies with different
kernel functions are shown in Fig.\,\ref{fig:In-every-test}. The
classification accuracy in most tests behaves as good as 100\%, which
demonstrates that the optimized models with linear, $3^{rd}$-order
polynomial, $4^{rd}$-order polynomial, and gaussian radial basis
kernel functions can all accurately sort out incorrect density data
of POINT system caused by common error sources.}{\large \par}

{\large{}}
{\large \par}

{\large{}For different kernel functions, the classification performances
evaluated by 24-fold cross-validation and corresponding model parameters
are listed in Table.\,\ref{tab:For-various-kernel}. The performance
of linear kernel function is outstandingly more accurate and stable
than other kernel functions. On the other hand, linear kernel function
has less kernel parameters to be optimized and faster training speed
\cite{hsu2003practical}. Therefore, the linear kernel function is
a good choice for data cleaning in this problem. With optimized parameters,
i.e., linear kernel function with penalty factor equals 8, and window
size of median filter equals 1000, the performance of TDGS method evaluated by 24-fold
cross-validation in cleaning density data of POINT system has reached 0.9871\textpm 0.0385. The average predication time of linear model for a new discharge is about 0.6752 $ms$. Moreover, the recall rate and false alarm of linear model for dissimilar samples are 0.9792 and 0.0294 respectively. Here the recall rate of dissimilar samples means the ratio of correctly predicted dissimilar samples to all dissimilar samples. The false alarm of dissimilar samples means the ratio of similar samples that are incorrectly predicted as dissimilar to all samples that are predicted as dissimilar. To evaluate this data cleaning model with a large set of discharges, we have picked out density data of POINT system under other 167 shots as test set. The performance of the optimized linear model in cleaning this new dataset has reached 0.9518 \textpm 0.0810. These data demonstrate that the optimized linear model
can be used as the final data cleaning model for real density data
analysis of POINT system reliably.}{\large \par}

{\large{}}
\begin{table}[h]
{\large{}\caption{For various kernel functions, the optimized classification performances
evaluated by 24-fold cross-validation and corresponding model parameters
in applying TDGS method to the density data cleaning of POINT system.
\label{tab:For-various-kernel} }
}{\large \par}
\centering{}{\large{}}%
\begin{tabular}{|c|c|c|c|}
\hline 
{\large{}Kernel functions} & {\large{}Performance} & {\large{}Penalty factor} & {\large{}Window size of median filter}\tabularnewline
\hline 
\hline 
{\large{}Linear} & {\large{}0.9871\textpm 0.0385} & {\large{}8} & {\large{}1000}\tabularnewline
\hline 
{\large{}Polynomial(3)} & {\large{}0.9712\textpm 0.0741} & {\large{}1} & {\large{}800}\tabularnewline
\hline 
{\large{}Polynomial(4)} & {\large{}0.9288\textpm 0.1178} & {\large{}5} & {\large{}2000}\tabularnewline
\hline 
{\large{}Rbf} & {\large{}0.9652\textpm 0.0597} & {\large{}1} & {\large{}1900}\tabularnewline
\hline 
\end{tabular}{\large \par}
\end{table}
{\large \par}

\begin{figure}[h]
{\large{}\caption{The absolute value of weight factors in the optimized linear classification
model for the density data cleaning of POINT system.\label{fig:The-absolute-value}}
}{\large \par}
\centering{}{\large{}\includegraphics[scale=0.4]{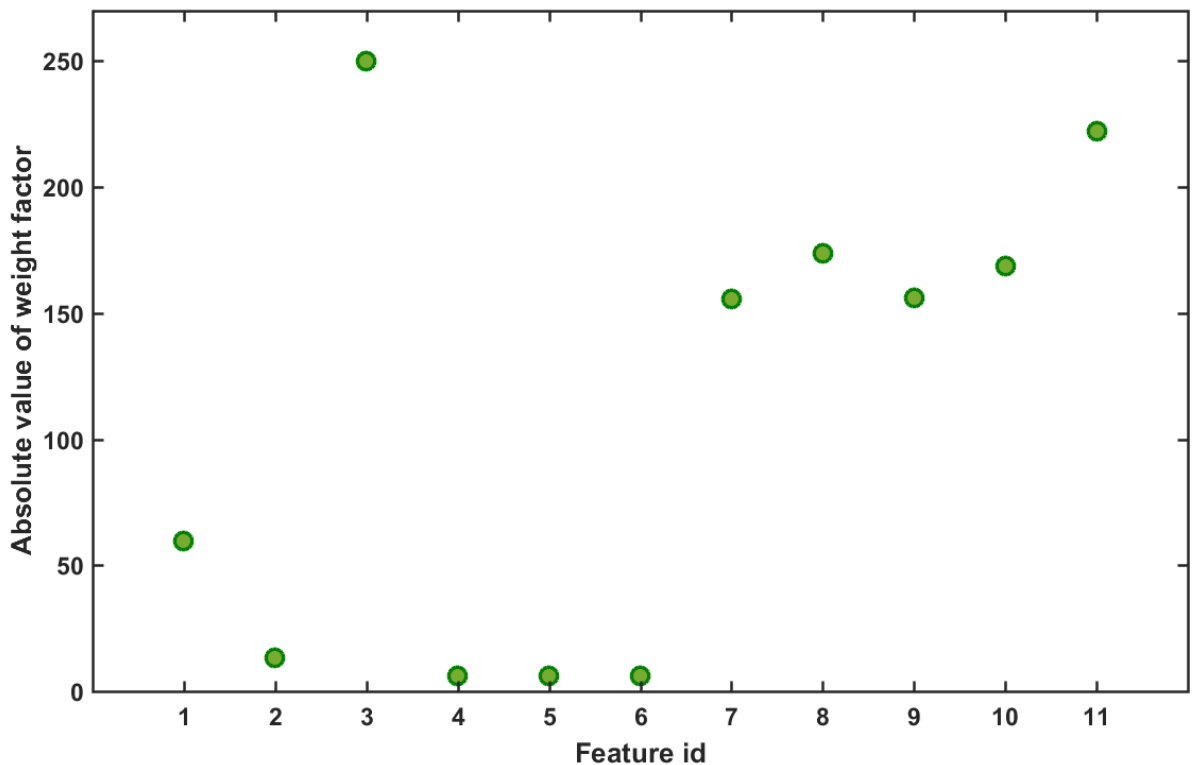}}{\large \par}
\end{figure}

{\large{}In linear SVM, the contributions of each features can be
ranked according to the absolute value of weight vector $\boldsymbol{W}$.
The larger $\left|W_{i}\right|$ is, the more important role is played
by the $ith$ feature in the linear model \cite{guyon2002gene}. The
absolute values of weight vectors in the final classification model
are shown in Fig.\,\ref{fig:The-absolute-value}. It can be observed
that each feature has evident contribution in this model, and the
contribution of feature 3, i.e., the Chebyshev Distance $\Vert\boldsymbol{S_{1}-S_{2}}\Vert_{\infty}$,
is the biggest. }{\large \par}

{\large{}}
{\large \par}

\section{{\large{}Summary}}

{\large{}In this paper, machine learning is applied to the automatic
data cleaning of MCF devices for the first time. Correct diagnostic
data sequences from different channels of MUM systems, which reflect
related yet distinct aspects of the same observed object, are physical
similar with each other. Based on this physical similarity, we propose
a general-purposed TDGS method to sort out the dirty diagnostic data
of MUM systems in MCF devices. The optimized performance of TDGS method evaluated by 24-fold
cross-validation in sorting the density data of POINT system has reached 0.9871\textpm 0.0385.}{\large \par}

{\large{}In the future, we will apply TDGS method to clean the POINT
data and other MUM systems in MCF devices. In large-scale applications
of TDGS method, predication speed and robustness should be further
considered as assessment indicators for model selection. Meanwhile,
the algorithm for data cleaning based on physical similarity between
data sequences should be further optimized to achieve less calculation
amount and error accumulation. Moreover, the physical similarity in
frequency-domain can also be utilized in some data cleaning problem
of MUM systems. }{\large \par}
\begin{acknowledgments}
This research is supported by Key Research Program of Frontier Sciences
CAS (QYZDB-SSW-SYS004), National Natural Science Foundation of China
(NSFC-11575185,11575186), National Magnetic Confinement Fusion Energy
Research Project (2015GB111003,2014GB124005), JSPS-NRF-NSFC A3 Foresight
Program (NSFC-11261140328),and the GeoAlgorithmic Plasma Simulator
(GAPS) Project. 
\end{acknowledgments}

\bibliographystyle{apsrev}
\bibliography{reference}

\end{document}